\documentclass[twoside,superscriptaddress,twocolumn,a4paper,floatfix,pra,aps,longbibliography]{revtex4-1}
\usepackage[utf8x]{inputenc}
\usepackage[colorinlistoftodos, color=green!40, prependcaption]{todonotes}
\usepackage{amsthm}
\usepackage{mathtools}
\usepackage{physics}
\usepackage{amsfonts}
\usepackage{xcolor}
\usepackage{graphicx}
\graphicspath{{plots/}}
\usepackage[left=23mm,right=13mm,top=35mm,columnsep=15pt]{geometry} 
\usepackage{adjustbox}
\usepackage{placeins}
\usepackage[T1]{fontenc}
\usepackage{lipsum}
\usepackage{csquotes}
\usepackage{hyperref}

\begin{document}
\title{Hierarchy and robustness of multilevel two-time temporal quantum correlations}
\author{Dawid Maskalaniec}\email[]{dawid_maskalaniec@outlook.com}
\affiliation{Faculty of Physics, Adam Mickiewicz University,
PL-61-614 Pozna\'n, Poland}
\author{Karol Bartkiewicz} \email[]{bark@amu.edu.pl}
\affiliation{Faculty of Physics, Adam Mickiewicz University,
PL-61-614 Pozna\'n, Poland}
\affiliation{RCPTM, Joint Laboratory of Optics of Palacký University 
and Institute of Physics of Czech Academy of Sciences, 17. listopadu 12, 
771 46 Olomouc, Czech Republic}

\date{\today}

\begin{abstract}
Quantum steering refers to correlations that can be classified as intermediate between entanglement and Bell nonlocality. Every state exhibiting Bell nonlocality exhibits also quantum steering and every state exhibiting quantum steering is also entangled. In low dimensional cases similar hierarchical relations have been observed between the temporal counterparts of these correlations. Here, we study the hierarchy of such temporal correlations for a general multilevel quantum system. We demonstrate that the same hierarchy holds for two definitions of state over time. In order to compare different types of temporal correlations, we show that temporal counterparts of Bell nonlocality and entanglement can be quantified with a temporal nonlocality robustness and temporal entanglement robustness. Our numerical result reveal that in contrast to temporal steering, for temporal nonlocality to manifest itself we require the initial state not to be in a completely mixed state.
\end{abstract}


\maketitle
\section{Introduction}
"Spooky action at the distance" between particles was firstly noted in the famous Einstein-Podolsky-Rosen (EPR) argument against the completeness of the Copenhagen interpretation of quantum mechanics~\cite{PhysRev.47.777}. In response to the EPR work, Schrödinger described a peculiar phenomenon, where one party can \textit{steer} the other, but not conversely~\cite{Schrodinger1935}. Nonlocal correlations were firstly considered as unnatural, even though they do not allow communication. For three decades the EPR paradox remained almost a philosophical question. However, in 1964 Bell proposed a test of realistic theories~\cite{PhysicsPhysiqueFizika.1.195}, which resulted in a great progress in theoretical and experimental research on quantum nonlocality~\cite{RevModPhys.86.419}. The systematic research on spatial steering initiated by Wieseman \textit{et al.}~\cite{PhysRevLett.98.140402} lead numerous results regarding various aspects of this phenomenon and its relation to other types of quantum correlations \cite{RevModPhys.92.015001}.  Remarkably, we can distinguish at least three different types of quantum correlations, i.e., entanglement, Bell nonlocality, and EPR steering. What is more, there exists a hierarchical relation between them. That is, every Bell nonlocal state is steerable and every steerable state is entangled, but not vice versa~\cite{PhysRevLett.98.140402}. As it turns out this hierarchy can be tested experimentally~\cite{Jirakova2021}.

\par In 1985 Legget and Garg, assuming \textit{macroscopic realism} (MR) and \textit{noninvasive measurability}, proposed a Bell-like experiment to test the MR of a single system at two different times \cite{Leggett1985,Emary2013}. Thereafter, other tests of MR were proposed, e.g., Chen \textit{et al.}~\cite{PhysRevA.89.032112} introduced a temporal counterpart of EPR steering inequality, which was further studied in relation to quantum key distribution protocols in~\cite{bartkiewicz2016temporal} or non-Markovianity~\cite{chen2016quantifying}. Next to these theoretical investigations, temporal steering was also reported experimentally~\cite{bartkiewicz2016experimental}. Another type of temporal correlations is the quantum causality. Confidence of validity of definite causal order is fundamental in our understanding of physical reality. However, recent research on quantum phenomena indicates that causality cannot be treated as an axiom. For example, there can exist superpositions of causual structures~\cite{Brukner2014} or other couterintuitive phenomena~ \cite{MacLean2017,Oreshkov2012,Chiribella2013,fitzsimons2015quantum}. The degree of casuality can be quantified by a monotone of causality proposed in~ \cite{fitzsimons2015quantum}. Hierarchy of temporal correlations under the assumption of no-signaling in time (NSIT) for two-level systems was shown in~\cite{mal2016probing,PhysRevA.98.022104}. Similarly to the spacial case, temporal Bell nonlocality implies temporal steering, and temporal steering implies temporal counterpart of entanglement, i.e., the quantum causal effect.

\par  The goal of this paper is to investigate relations between temporal correlations in systems of arbitrary dimensions. For systems larger than qubits, the spooky action over time requires more study. Theese can exhibit unusual properties as it happens to be the case with studying quantum contextuality~\cite{Budroni2021}, which can be only studied in systems larger than qubits. To analyze such systems, we extend the notion of a pseudo density operator (PDO), originally introduced by Fitzsimons \textit{et al.}~\cite{fitzsimons2015quantum} for multi-qubit system, to describe odd-dimensional systems, and compare it to a Wigner function formulation from Ref.~\cite{WOOTTERS19871}. In Sec.~\ref{sec:results}, it is proven that the hierarchy between temporal nonlocality, temporal steering, and temporal entanglement holds under the NSIT assumption. However, as we also demonstrate here, in systems which do not satisfy the NSIT condition the hierarchy can be broken.

\par Quantification and description of the quantum correlations can be performed in various ways, including, e.g., a geometric approach used for EPR steering in Ref.~\cite{Ku2018} or matrices of moments applied for detecting various nonclassical correlations (see, e.g., Ref.~\cite{Miranowicz2010}). However, to treat all types of temporal correlations consistently, we decided to quantify robustness of quantum correlations. In Sec.~\ref{sec:framework} we introduce new quantifiers of robustness of quantum correlations, that is \textit{temporal nonlocality robustness} (TNR) and \textit{temporal entanglement robustness} (TER) as direct counterparts of spatial nonlocality robustness~\cite{PhysRevA.93.052112} and spacial entanglemnt robustness~\cite{PhysRevA.59.141}, respectively. The comparison of TNR, TER, and \textit{temporal steering robustness} (TSR) included in this paper provides a systematic picture of the correlations. Our concussions are summarized in Sec.~\ref{sec:conclusions}.

\section{Theoretical framework}\label{sec:framework}
\subsection{States over time and temporal entanglement}
The notion of quantum mechanical states defined over time is far from being a trivial counterpart of a single-time multidimensional state~\cite{PhysRevA.79.052110,PhysRevA.88.052130,fitzsimons2015quantum,Horsman2017}. The definition of a pseudo-density operator (PDO), that we extend in this paper, was introduced in Ref.~\cite{fitzsimons2015quantum} in order to establish a formalism which treats space and time indiscriminately. This proposition of state over time was further studied in \cite{Zhao2018,Pisarczyk2019,Zhang2020,Zhang_2020-2}.Analogously to $n$ qubits in space, let us consider $n$ events in space-time, where at each event a measurement of a single Pauli operator can be performed. A density matrix describing system composed of $n$ qubits can be written as
\begin{equation}
    \rho = \frac{1}{2^n}\sum_{i_1=0}^3\dots\sum_{i_n=0}^3\left\langle
    \bigotimes_{j=1}^n\sigma_{i_j}\right\rangle\bigotimes_{j=1}^n\sigma_{i_j}, \label{eq:nqubit_dm}
\end{equation}
where $\sigma_i$ are Pauli matrices. This leads to Fitzsimons' \textit{et al.} proposal for a state over time~\cite{fitzsimons2015quantum}:
\begin{equation}
    R^{\text{PDO}} = \frac{1}{2^n}\sum_{i_1=0}^3\dots\sum_{i_n=0}^3\left\langle\left\{\sigma_{i_j}\right\}_{j=1}^n\right\rangle\bigotimes_{j=1}^n\sigma_{i_j},\label{eq:R_PDO}
\end{equation}
where $\langle\{\sigma_{i_j}\}_{j=1}^n\rangle=\textrm{tr}[(\bigotimes_{j=1}^n\sigma_{i_j})R^{\text{PDO}}]$ is the expectation value of the product of the results of measurements. In spite of the fact that definition (\ref{eq:R_PDO}) for two instances refers to measurements performed at $t_A$ and $t_B$,  $R^{\text{PDO}}$ describes an initial state $\rho_A$ evolving to $\rho_B$ under the quantum channel $\mathcal{E}:\rho_A\mapsto\rho_B$, represented via Choi--Jamiołkowski isomorphism \cite{jamiolkowski1972linear,choi1975completely} as operator
\begin{equation}
    E_{B\vert A}=\sum_{i,j}\vert i\rangle\langle j\vert_A\otimes\mathcal{E}(\vert j\rangle\langle i\vert_A).\label{eq:E_BA}
\end{equation}
Density matrices $\rho_A$ and $\rho_B$ are spanned on $\mathcal{H}_A$ and $\mathcal{H}_B,$ respectively. Hence, the PDO is explicitly determined by the initial state and the channel.

\par However, a state over time can be defined in more than one way. For example, in Ref.~\cite{WOOTTERS19871} authors construct it using the quasi-probabilities from a discrete Wigner representation \cite{Gibbons2004,Gross2006}, which is defined by a set of phase-space point operators $\{K_i\}_{i=0}^{d^2-1}$. Operators $K_i$ form a basis for a $d^2$-dimensional Hilbert space $\mathcal{H}$ with the Hilbert-Schmidt product $\langle X\vert Y\rangle=\tr(XY)$. They satisfy $\sum_{i}K_i=d\openone,$ $\tr(K_i K_j)=d\delta_{ij}$ and $\tr(K_i)=1$. Since arbitrary density matrix $\rho_A$ and operator $E_{B\vert A}$ can be written as $\sum_{i}r(i)K_i^A$ and $\sum_{i,j}r(j\vert i)K_i^A\otimes K_j^B$, respectively, one may describe an evolution of a quantum system between events $A$ and $B$ by a different operator
\begin{equation}
    R^{\text{W}} = \sum_{i,j}r(j\vert i)r(i)K_{i}^{A}\otimes K_{j}^B,\label{eq:R_W}
\end{equation}
where quasi-probability distributions $r(i)$ and $r(j\vert i)$ are given by $\tr(\rho_A K_i)$ and $\tr[E_{B\vert A}(K_i^A\otimes K_j^B)]$, respectively. In general, the properties of various operators $R$ can be studied via the star product $\star$ (see Ref.~\cite{Horsman2017}), a binary operation $\star:X\times X\mapsto X\in\mathcal{H}_A\otimes\mathcal{H}_B$.
In this framework a whole class of definitions of a sate over time 
\begin{equation}
    R = \rho_A\star E_{B\vert A}\label{eq:star}
\end{equation}
can be represented regardless of their explicit form.

\par A given state over time does not need to be positive semi-definite. 
If $R\geq 0$, then it might be considered as a regular density matrix. However, negative eigenvalues of the state over time imply time-like correlations between events. A proper causality measure for these events $\Phi(R)$ has to distinguish between space-like and time-like correlations and be a causality monotone. A given function $\Phi(R)$ is a causality monotone if \cite{fitzsimons2015quantum}:
\begin{itemize}
    \item $\Phi(R)\geq 0$, with $\Phi(R)=0$ if $R$ is completely positive, 
    and $f(R)$ is maximal for any $R$ obtained from two consecutive measurements on a closed system,
    \item $\Phi(R)$ is invariant under unitary operations,
    \item $\Phi(R)$ is non-increasing under local operations,
    \item $\sum_ip_i\Phi(R_i)\geq \Phi(\sum_ip_iR_i)$.
\end{itemize}
The above criteria are similar to these for an entanglement monotone \cite{RevModPhys.81.865}. Therefore, we refer to the phenomena for which $\Phi(R)\neq 0$ as \textit{temporal entanglement}. In Ref.~\cite{fitzsimons2015quantum} authors proposed a simple monotone, i.e., $f(R)=\|R\|_{\tr}-1$. A separable state over time can be expressed as
\begin{equation}
    R^{\mathrm{SEP}}=\sum_k p_k\rho_k^A\otimes\rho_k^B,\label{eq:separableR}
\end{equation}
where $p_k$ stands for a certain probability distribution, $\rho_k^A$, $\rho_k^B$ are density matrices spanned on Hilbert spaces $\mathcal{H}_A$ and $\mathcal{H}_B,$ respectively. This can be interpreted as a space-like PDO being generated from a maximally entangled space-like state, where subsystem $B$ is subjected to a dissipative channel. However, $R\geq 0$ does not imply separability, i.e., there can still exist space-like temporal correlations. A quantum state is \textit{spatially entangled,} if it cannot be written as a convex combination of product states (\ref{eq:separableR}). Otherwise the state is \textit{separable}. 

\par Quantum entanglement of a given state $\sigma$ can be quantified, e.g., as the minimal amount of noise $\tau$ that makes the state separable~\cite{PhysRevA.59.141}. This entanglement robustness (ER) can be expressed as
\begin{eqnarray}
    \text{ER}=&&\,\min\,\gamma,\nonumber\\
    \text{s.t. }&& \frac{\sigma+\gamma\tau}{1+\gamma}=\rho,\nonumber\\
    &&\rho=\sum_k p_k\rho_k^A\otimes\rho_k^B,\nonumber\\
    &&p_k\geq 0,\quad\rho_k^A,\rho_k^B\geq 0\quad\forall k,\nonumber\\
    &&\tr(\tau)=\tr(\rho_k^A)=\tr(\rho_k^B)=1\quad\forall k,\nonumber\\
    &&\sum_kp_k=1,\quad\tau\geq 0,\quad\gamma\geq 0.\label{eq:ER}
\end{eqnarray}
By analogy, we introduce temporal entanglement robustness (TER) as minimal amount of noise $\mathfrak{R}$ needed to destroy temporal causality. The relevant optimization problem reads
\begin{eqnarray}
    \text{TER}=&&\,\min\,\gamma\nonumber\\*
    \text{s.t. }&&\frac{R+\gamma\mathfrak{R}}{1+\gamma}=\rho,\nonumber\\*
    &&\mathfrak{R}:\;\text{a pseudo-density operator},\nonumber\\*
    &&\rho\geq 0,\quad\gamma\geq 0.\label{eq:TER}
\end{eqnarray}
By substituting the product $\gamma\mathfrak{R}$ with unnormalized state over time $\tilde{\mathfrak{R}}$ we can easily rewrite the above problem in a linear SDP form,
\begin{eqnarray}
    &&\text{TER}=\min\left(\tr\tilde{\mathfrak{R}}\right),\nonumber\\*
    &&\text{s.t. }R+\tilde{\mathfrak{R}}\geq 0.\label{eq:TER2}
\end{eqnarray}
We shown in Appendix\ref{App:A} that TER (\ref{eq:TER}) satisfies the above-listed criteria for a causality monotone.

\subsection{Temporal steering}
Consider two parties $A$ (Alice) and $B$ (Bob). At time $t_A$ Alice receives an initial state $\rho_A$. Then, she performs a positive-operator valued measurement (POVM) $\{M_{a\vert x}\}$ and obtains result $a$ with probability $p(a\vert x)=\tr_A(\rho_A M_{a\vert x})$. Next, the state
\begin{equation}
    \rho_{a\vert x}\left(0\right)=\sqrt{M_{a\vert x}}\rho_A\sqrt{M_{a\vert x}}\label{eq:init_conditional_state}
\end{equation}
is delivered to $B$ through a channel represented by a Choi-Jamiołkowski operator $E_{B\vert A}$. Bob receives the conditional state
\begin{eqnarray}
    \rho_{a\vert x}\left(t\right)&&\equiv\rho_{a\vert x}=\mathcal{E}\left(\sqrt{M_{a\vert x}}\rho_A\sqrt{M_{a\vert x}}\right)\nonumber\\*
    &&{}=\tr_A\left(E_{B\vert A}\sqrt{M_{a\vert x}}\rho_A\sqrt{M_{a\vert x}}\right).\label{eq:conditional_state}
\end{eqnarray}
If Alice and Bob repeat their protocol many times, they can reconstruct their shared set of unnormalized conditional states $\{\tilde{\rho}_{a\vert x}=p(a\vert x)\rho_{a\vert x}\},$ referred to as \textit{assemblage}. 

\par If Bob wants to explain receiving $\tilde{\rho}_{a\vert x}$ with a \textit{hidden state model} (HSM), he can assume that Alice sends state $\tilde{\rho}_{a\vert x}$ through different channels according to probability distribution $p(\lambda)$. Hence, there exists a classical variable $\lambda$ which determines both the Alice's outcome and the conditional state $\rho_{a\vert x}=\rho_\lambda$. Thus, according to Bob, a conditional state can be written as
\begin{equation}
    \rho_{a\vert x}=\sum_\lambda p(\lambda)D(a\vert x,\lambda)\rho_\lambda\quad\forall a,x.\label{eq:HSM}
\end{equation}
If the state $\rho_{a\vert x}$ can be expressed as in Eq.~(\ref{eq:HSM}), Bob concludes that the state at time $t_A$ does not influence the assemblage $\rho_{a\vert x}$. Thus, he excludes the possibility of Alice steering his state. In such case $\rho_{a\vert x}$ is \textit{temporally unsteerable}. Conversely, if a conditional states do not admit the decomposition (\ref{eq:HSM}), then Bob concludes that Alice steers his state. We refer to assemblages which do not admit the HSM as \textit{temporal steerable}. 

\par In this paper we quantify the steerability of an assemblage with \textit{temporal steering robustness} (TSR), introduced in \cite{PhysRevA.94.062126,Chen2017SciRep} as a temporal couterpart of steering robustness \cite{PhysRevLett.114.060404}. TSR is defined as the minimal amount of noise needed to destroy the temporal steerability of a given assemblage:
\begin{eqnarray}
    \text{TSR}=&&\,\min\,\alpha\nonumber\\*
    \text{s.t. }&&\left\{\frac{\tilde{\rho}_{a\vert x}+\alpha\tau_{a\vert x}}{1+\alpha}\right\}_{a,x}\text{ temporal unsteereable,}\nonumber\\*
    &&\left\{\tau_{a\vert x}\right\}:\,\text{a noisy assemblage.}\label{eq:TSR}
\end{eqnarray}
The optimization problem (\ref{eq:TSR}) can be rewritten as a semi-definite program (SDP)
\begin{eqnarray}
    \text{TSR}=&&\min\left(\tr\sum_\lambda\tilde{\rho}_\lambda-1\right)\nonumber\\*
    \text{s.t. }&&\sum_\lambda D(a\vert x,\lambda)\tilde{\rho}_\lambda-\rho_{a\vert x}\geq0\quad\forall a,x,\nonumber\\*
    &&\tilde{\rho}_\lambda\geq0\quad\forall\lambda,\label{eq:TSR_SDP}
\end{eqnarray}
where $\tilde{\rho}\equiv(1+\alpha)\rho_\lambda$ and $D(a\vert x,\lambda)=\delta_{a,\lambda(x)}.$

\subsection{Temporal counterpart of Bell nonlocality}
Another type of correlations, different than steering, can be detected by a Bell-type experiment. Let us consider a scenario, where Alice and Bob make measurements of observables $A_x$ and $B_y,$ respectively. Quantum mechanics predicts that the probability of obtaining results $a$ and $b$ is given by the Born rule, i.e.,
\begin{eqnarray}
    P(a,b\vert x,y)=\tr\left[M_{b\vert y}\mathcal{E}\left(\sqrt{M_{a\vert x}}\rho_A\sqrt{M_{a\vert x}}\right)\right]\nonumber\\
    \forall a,b,x,y,\label{eq:p(ab|xy)}
\end{eqnarray}
where $M_{c\vert z}$ for $c=a,b$ and $z=x,y$ stands for a POVM describing the measurement of value $c$ associated with observable $C_z$ (i.e. $A_z$ or $B_z$). The joint probability distribution $\{P(a,b\vert x,y)\}_{a,b,x,y}$ is also referred to as \textit{behavior} of the system.

\par A \textit{local hidden variable} (LHV) model can be constructed, if the behavior $P(a,b\vert x,y)$ can be expressed as
\begin{equation}
    P(a,b\vert x,y)=\sum_{\mu,\nu}p(\mu,\nu)D(a\vert x,\mu)D(b\vert y,\nu),\label{eq:LHV}
\end{equation}
where $\mu$ and $\nu$ are classical random variables given by a preexisting distribution $p(\mu,\nu),$ and $D(a\vert x,\mu)$ and $D(b\vert y,\nu)$ are local deterministic response functions. The existence of an LHV model implies that the behavior $P(a,b\vert x,y)$ is macro-realistic and that correlations between $A$ and $B$ can origin in a third party controlling measurement devices at $A$ and $B$.

\par We refer to the sets of probability distributions which cannot be expressed by Eq.~(\ref{eq:LHV}) as \textit{nonlocal}. We quantify the degree of Bell-nonlocality with a temporal version of nonlocality robustness from Ref.~\cite{PhysRevA.93.052112}. Temporal nonlocality robustness (TNR) is defined as the minimal amount of behavior $Q(a,b\vert x,y)$ needed to make $P(a,b\vert x,y)$ local,
\begin{eqnarray}
        \text{TNR}&&{}=\min\,\beta,\nonumber\\*
        \text{s.t. }&&\left\{\frac{P(a,b\vert x,y)+\beta Q(a,b\vert x,y)}{1+\beta}=R(a,b\vert x,y)\right\},\nonumber\\*
        &&R(a,b\vert x,y)=\sum_{\mu,\nu}r(\mu,\nu)D(a\vert x,\mu)D(b\vert y,\nu),\nonumber\\*
        &&\left\{Q(a,b\vert x,y)\right\}\text{: a behavior,}\quad\beta\geq 0.\label{eq:TNR}
\end{eqnarray}
Additionally, we can introduce local hidden variable temporal nonlocality robustness (LHV TNR) by requiting that the noise is an LHV behavior, i.e.,
\begin{eqnarray}
    Q(a,b\vert x,y)=&&{}\sum_{\mu,\nu}q(\mu,\nu)D(a\vert x,\mu)D(b\vert y,\nu),\nonumber\\*
    &&q(\mu,\nu)\geq 0\quad\forall\mu,\nu.
\end{eqnarray}
Following the techniques presented in \cite{PhysRevLett.114.060404}, one may rewrite this problem in the linear SDP form
\begin{eqnarray}
     \text{TNR}&&{}=\min\left[\frac{\sum_{x,y,\mu,\nu}\tilde{r}(\mu,\nu)}{\sum_{x,y,a,b}P(a,b\vert x,y)}-1\right],\nonumber\\*
     \text{s.t. }&&\sum_{\mu,\nu}\tilde{r}(\mu,\nu)D(a\vert x,\mu)D(b\vert y,\nu)\geq P(a,b\vert x,y),\nonumber\\*
     &&\tilde{r}(\mu,\nu)\geq 0\quad\forall\mu,\nu,a,b,x,y,\label{eq:TNR_SDP}
\end{eqnarray}
where $\tilde{r}(\mu,\nu)$ stands for the unnormalized probability distribution $(1+\beta)r(\mu,\nu)$. For details of the derivation see Appendix~\ref{App:B}. In the case of LHV TNR we require that there exists such $\tilde{q}(\mu,\nu)$ that 
\begin{eqnarray}
    \sum_{\mu,\nu}\tilde{r}&&(\mu,\nu)D(a\vert x,\mu)D(b\vert y,\nu)\nonumber\\*
    =&&{}\sum_{\mu,\nu}\tilde{r}(\mu,\nu)D(a\vert x,\mu)D(b\vert y,\nu)-P(a,b\vert x,y),\nonumber\\*
    &&{}\tilde{q}(\mu,\nu)\geq 0,\quad\forall a,b,x,y.
\end{eqnarray}

\par Both temporal Bell inequalities and LHS model assume non-invasive measurements, for which a necessary condition is the no-signaling in time (NSIT) condition \cite{Kofler2013,Halliwell2017}. NSIT is obeyed if the Alice's measurement does not change the outcome statistics of the Bob's measurement. It implies that the elements of assemblage have to satisfy the following relation,
\begin{equation}
\sum_a\varrho_{a\vert x}=\tr_A\left(E_{B\vert A}\rho_A\right)\quad\forall x.\label{eq:NSIT}
\end{equation}
To demonstrate that a state does not satisfy the NSIT condition it is sufficient to find such a pair of $x_1,x_2$ for which
\begin{equation}
\sum_a\varrho_{a\vert x_1}\neq\sum_a\varrho_{a\vert x_2}.
\end{equation}
This is the case if we choose the initial state to be a pure state, i.e., an eigenstate of $x_1$ and the channel is the identity channel. Then the eigenvectors of operator $x_2$ form mutually unbiased set with respect to the eigenvectors of $x_1.$ In this extreme case, one side of the equality corresponds to a pure state, while the other to a maximally mixed state. Thus, we can expect pure states to violate the NSIT condition for some channels.

\section{Results}\label{sec:results}
\subsection{Bipartite pseudo-density operator of a \textit{d}-dimensional system}
Let us consider a quantum system AB composed of two qudits. Assume that operators $G_i^A$ and $G_j^B$ form orthogonal basis (i.e., $\tr\left(G_iG_j\right)=d\delta_{i,j},$ where $d>0$) in $\mathcal{H}_A$ and $\mathcal{H}_B$, respectively. A density matrix describing AB can be written in terms of $G_i^A$ and $G_j^B$ as
\begin{equation}
    \rho_{AB}=\frac{1}{\mathcal{N}}\sum_{i=0}^{d_A^2-1}\sum_{j=0}^{d_B^2-1}C_{ij}G_i^A\otimes G_j^B,\label{eq:rho1}
\end{equation}
where $\mathcal{N}$ is a normalization factor such that $\tr(\rho_{AB})=1$ and $C_{ij}=\mathcal{N}\tr[\rho_{AB}(G_i^A\otimes G_j^B)]$ is a correlation tensor. To make the following expressions more legible, from this point we will denote all operators simply as $G_i$ and assume that $d_A = d_B=d$.

\par Equation (\ref{eq:rho1}) suggests the following generalization of the notion of the PDO,
\begin{equation}
    R^{\text{PDO}}=\frac{1}{\mathcal{N}}\sum_{i,j=0}^{d^2-1}\left\langle G_i\otimes G_j\right\rangle G_i\otimes G_j,\label{eq:new_PDO}
\end{equation}
where $\langle G_i\otimes G_j\rangle$ stands for the expectation value of the product of the result of measurements of operators $G_i$ and $G_j$ at $t_A$ and $t_B$, respectively. Formally, this reads
\begin{eqnarray}
    \left\langle G_i\otimes G_j\right\rangle&=&\tr\left[R^{\text{PDO}}\left(G_i\otimes G_j\right)\right]\nonumber\\*
    &=&\sum_a a\tr\left[E_{B\vert A}\left(\rho_{i,a}^A\otimes G_j\right)\right].\label{eq:correlation_tensor}
\end{eqnarray}
Operator $\rho_{i,a}^A=\Pi_{i,a}\rho_A\Pi_{i,a}$ denotes the subnormalized state of $A$ 
after projection $\Pi_{i,a}$ onto the eigenspace corresponding to an eigenvalue $a$ of operator $G_i.$

\par Since in 2D case the PDO is defined by the use of Pauli matrices, it would be natural to establish $G_i$ as generators of special unitary group $\text{SU}(d)$. However, due to contextuality~\cite{Budroni2021} between measurements of the $\text{SU}(d)$ generators, such definition is dependent on a choice of basis of $\mathcal{H}_A$ and $\mathcal{H}_B$. One may explicitly verify that $R^{\text{PDO}}$ has different spectrum for different pure initial states subjected to the identity channel. Entanglement is not base-dependent so we cannot quantify it reliably utilizing base-dependent quantities. Therefore, we have to use non-contextual scenarios to define a proper state over time. For systems of odd, prime dimension the subset of quantum states known to be classically simulable, and thus forming non-contextual scenario, is prescribed by the Wigner polytope \cite{Gross2006,Dawkins2015PRL},
\begin{equation}
    \text{Wigner polytope}=\{\rho\,\vert\,\tr(\rho K_i)\geq 0\;\forall i\}.
\end{equation}
Hence, by setting $G_i=K_i$ we make $R^{\text{PDO}}$ independent of a chosen basis for quantum systems being in the Wigner polytope. It is worth emphasizing that definitions (\ref{eq:R_W}) and (\ref{eq:new_PDO}) are not equivalent. Quasi-probability distribution $r(j\vert i)r(i)$ differs from expectation values $\langle G_i\otimes G_j\rangle$ used in equations (\ref{eq:R_PDO}) and (\ref{eq:new_PDO}).

\par In particular, for a qutrit (see Ref~\cite{Dawkins2015PRL}) the discrete Wigner-space point operators $K_i$ operators can be selected as 
\begin{equation*}
K_1=\left[
\begin{array}{ccc}
 1 & 0 & 0 \\
 0 & 0 & 1 \\
 0 & 1 & 0 \\
\end{array}\right],\,
K_2=\left[
\begin{array}{ccc}
 0 & 1 & 0 \\
 1 & 0 & 0 \\
 0 & 0 & 1 \\
\end{array}\right],\,
K_3=\left[
\begin{array}{ccc}
 0 & 0 & 1 \\
 0 & 1 & 0 \\
 1 & 0 & 0 \\
\end{array}\right],
\end{equation*}
\begin{eqnarray*}
K_4&=&\left[
\begin{array}{ccc}
 1 & 0 & 0 \\
 0 & 0 & \frac{-1-\sqrt{3}i}{2} \\
 0 &  \frac{-1+\sqrt{3}i}{2} & 0 \\
\end{array}\right],\\
K_5&=&\left[
\begin{array}{ccc}
 0 &  \frac{-1-\sqrt{3}i}{2} & 0 \\
  \frac{-1+\sqrt{3}i}{2} & 0 & 0 \\
 0 & 0 & 1 \\
\end{array}\right],\\
K_6&=&\left[
\begin{array}{ccc}
 0 & 0 & \frac{-1-\sqrt{3}i}{2} \\
 0 & 1 & 0 \\
 \frac{-1+\sqrt{3}i}{2} & 0 & 0 \\
\end{array}\right],\\
\end{eqnarray*}
\begin{equation}
K_7 = K_4^T,\qquad
K_8 = K_5^T,\qquad
K_9 = K_6^T.
\end{equation}
These matrices differ from matrices usually used to express a density matrix of three-level system, i.e., 
Patera-Zassenhaus~\cite{Patera1988JMP} matrices, Gell-Mann~\cite{Gell-Mann1962PR} matrices, or matrices of spin-1 spherical symmetries~\cite{Hofmann2004PRA}. We confirmed numerically our prediction that applying matrices $K$ in the definition of a PDO for qutrits makes the spectrum of a PDO independent of a particular choice the pure initial state.

\subsection{Relations between temporal entanglement and separability for qubits and three-level systems}
 
\begin{figure*}
\includegraphics[width=1\linewidth]{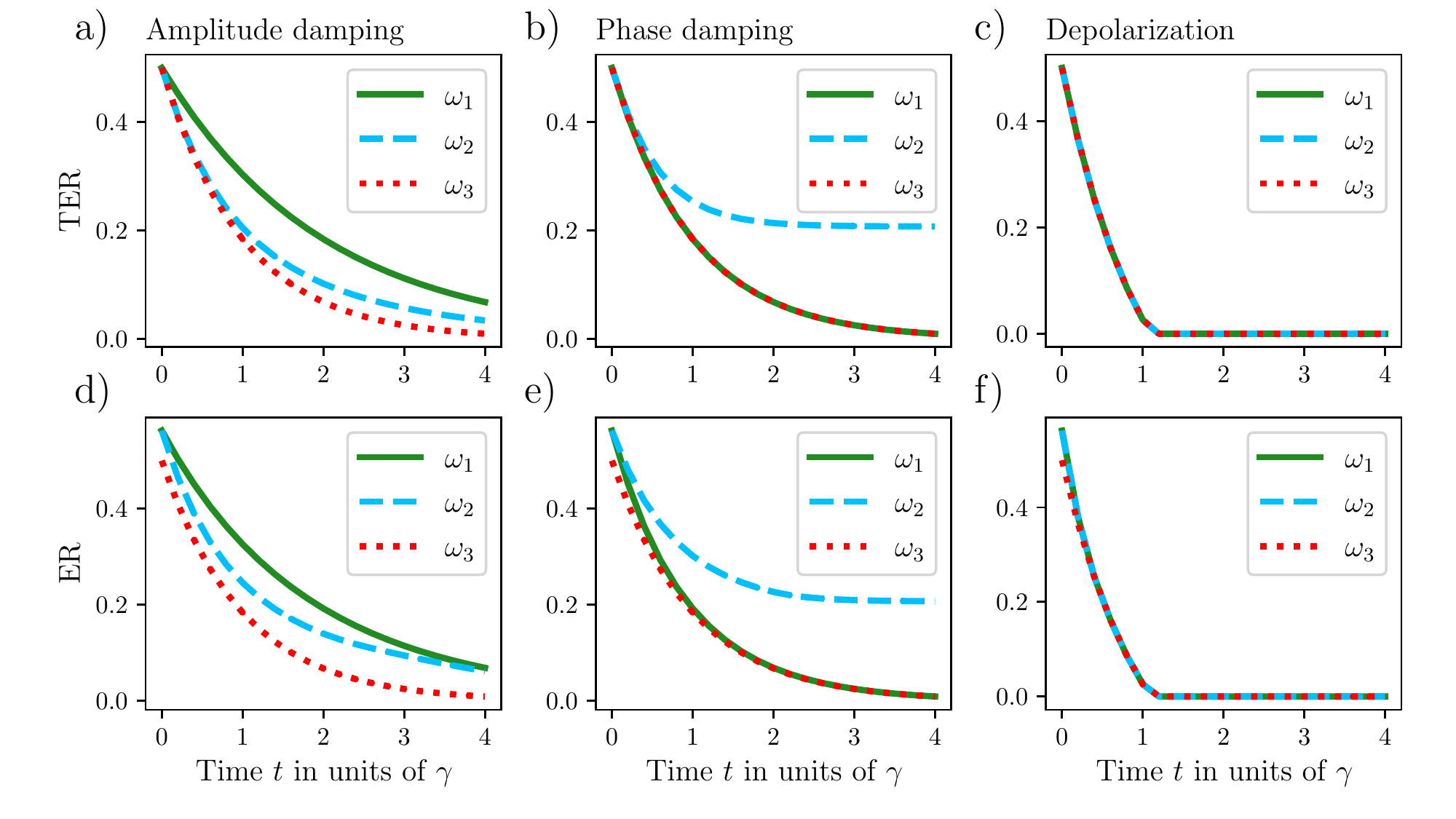}
  \caption{(color online) The dynamics of (a)-(c) TER  and  (d)-(f) ER for a qubit initially in state evolving under [(a),(d)] amplitude damping channel, [(b),(e)] phase damping channel, and [(c),(f)] depolarizing channel.  Time $t$ is measured in units of decay rate $\gamma$. Initially the system is in: (solid curve) vacuum state $\omega_1:\vert 0\rangle,$, (dashed curve) balanced superposition $\omega_2:\frac{1}{\sqrt{2}}(\vert 0\rangle + \vert 1\rangle),$ and (dotted curve) maximally mixed state $\omega_3:\openone/2.$ We observe that for systems initially in pure states positive semi-definiteness of PDO does not imply separability. For depolarization we observe vanishing of TER and ER for the same finite time. Also ER calculated for $\omega_3$ (in contrast to pure states) is above $1/2,$ which is an idicator of a physical inequivalence between TER and ER.}
  \label{fig:ERvsTER}
\end{figure*}

In Ref.~\cite{PhysRevA.98.022104} the authors proved that a PDO describing a maximally mixed qubit transmitted through amplitude damping, phase damping or depolarizing channel is separable, if $R^{\text{PDO}}$ is positive semi-definite. Let us firstly investigate the case of a pure initial state. In the cases of qubit-qubit and qubit-qutrit bipartite systems Peres-Horodecki positive partial transpose (PPT) criterion \cite{PhysRevLett.77.1413,HORODECKI19961} is both necessary and sufficient condition for separability of the system. This criterion can be applied to simplify the optimization problem (\ref{eq:ER}) to a linear SDP form, i.e.,
\begin{eqnarray}
    \text{ER}=&&{}\min\left[\tr\left(\mathfrak{R}\right)-1\right],\nonumber\\*
    \text{s.t. }&&R+\mathfrak{R}=\rho,\nonumber\\*
    &&\rho\geq 0,\nonumber\\*
    &&\rho^{\mathrm{PT}}\geq 0,\label{eq:ER-2level}
\end{eqnarray}
where $\mathrm{PT}$ stands for a partial transposition. By considering a simple example of a qubit in a pure state at time $t_A$ subjected to amplitude damping, phase damping or depolarizing channel we can demonstrate that TER (\ref{eq:TER}) and ER (\ref{eq:ER-2level}) are not equivalent (see Fig.~\ref{fig:ERvsTER}). It is clear that space-like separable state over time is also temporally separable.

\par In the case of a system of an arbitrary dimension the sufficient condition is that for all positive maps $\Lambda$ acting on $\mathcal{H}$, a separable density matrix of a composite system $\rho\in\mathcal{H}\otimes\mathcal{H}$ satisfies $(\Lambda\otimes\openone)\rho\geq 0$. Equivalently, one can find an entanglement witness $W\in\mathcal{H}\otimes\mathcal{H}$, Choi-isomorphic to $\Lambda$, such that $\tr(W\rho)\geq 0$ necessarily for a separable matrix $\rho$ and $\tr(W\rho)\leq 0,$ otherwise~\cite{GUHNE20091,RevModPhys.81.865}. However, testing $\rho$ under all of positive maps $\Lambda$ is impractical. Hence, we are forced to use necessary but not sufficient criteria \cite{GUHNE20091,RevModPhys.81.865} in order to detect spacial entanglement. These criteria are not universal but they can be easily applied and allow to distinguish many separable states. In Appendix~\ref{App:C} we show that linear separability criteria based on correlation matrix, which have been recently unified in Ref.~\cite{PhysRevA.101.012341}, are satisfied by a positive semi-definite PDO representing a single qutrit, initially in maximally mixed state, undergoing depolarizing, phase damping, and amplitude damping. Hence, it is reasonable to assume, as in Ref.~\cite{PhysRevA.98.022104} for the two-level case, that
\begin{eqnarray}
    &&R\geq 0\,\Rightarrow\,\exists\,\rho_k^A,\,\rho_k^B,\nonumber\\*
    &&\text{such that }R=\sum_kp_k\rho_k^A\otimes\rho_k^B,\,\rho_A=\tfrac{1}{3}\openone.\label{eq:proposition}
\end{eqnarray}
{However, proving this proposition is beyond the scope of this paper.}

\subsection{Hierarchy of two-time three-level correlations}

\begin{figure*}
\includegraphics[width=\linewidth]{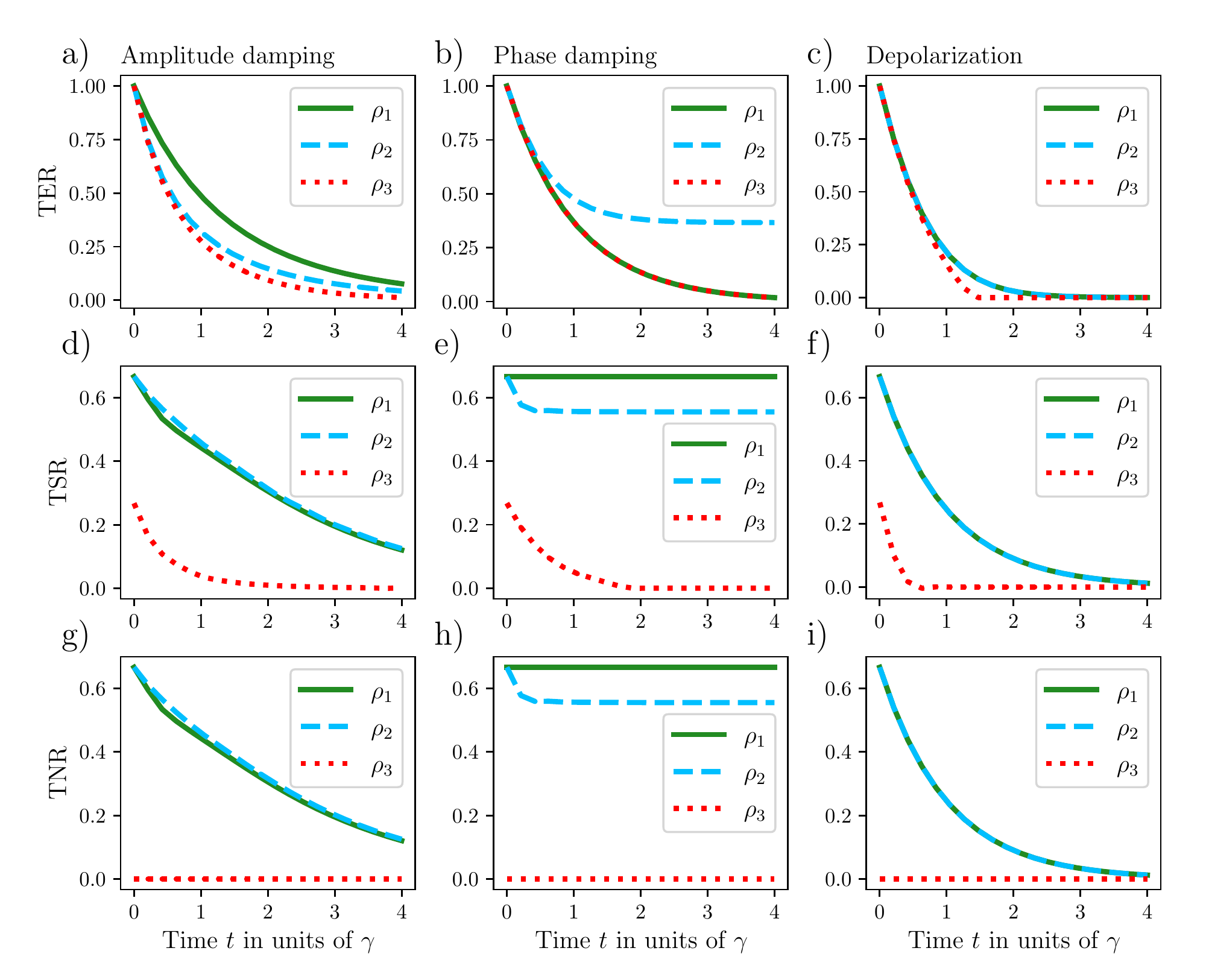}
\caption{(color online) The dynamics of (a)-(c) TER, (d)-(f) TSR, and (g)-(i) TNR for a qutrit evolving under [(a),(d),(g)] amplitude damping channel, [(b),(e),(h)] phase damping channel, [(c),(f),(i)] depolarizing channel. Time $t$ is measured in units of decay rate $\gamma$. Initially the system is in (solid curve) vacuum state $\rho_1:\ket{0}$ state, balanced superposition (dashed curve) $\rho_2:(\ket{0}+\ket{1}+\ket{2})/\sqrt{3},$ or (dotted curve) maximally mixed state $\rho_3:\openone/3.$ We can see that in all cases $\text{TSR}\geq\text{TNR}$, which demonstrates the hierarchical relation between temporal steering and nonlocality. If the initial state satisfies NSIT condition the standard hierarchy of correlations is conserved, $\text{TER}\geq\text{TSR}$. However, panels (c) and (f) demonstrate that in general even a positive semi-definite state over time can describe a temporal steerable system. Note that in all cases involving depolarizing channel we can observe the sudden death of temporal entanglement.}
\label{fig:hierarchy_pdm}
\end{figure*}

\par Hierarchical relations between two-time correlations in systems of arbitrary dimension, obeying NSIT condition, can be proven in the same manner as in two-level case \cite{PhysRevA.98.022104}. Let us start with a state that satisfies Eq.~(\ref{eq:NSIT}), i.e., $\rho_A=\openone/d$. To show the hierarchy, let us recall the scenario demonstrating temporal steering. At time $t_A$ Alice performs a measurement represented by a set of POVMs $\{M_{a\vert x}\}_{x,a}$ on a state $\rho_A$ and sends to Bob a temporal assemblage $\{\rho_{a\vert x}\}$ given by Eq.~(\ref{eq:conditional_state}). It is easy to see that for maximally mixed initial states both formulations of a state over time (\ref{eq:R_W}, \ref{eq:new_PDO}) reduce to the Choi-Jamiolkowski operator (\ref{eq:E_BA}). Hence, the assemblage can be written in terms of a state over time $R_{AB}$ as
\begin{equation}
    \rho_{a\vert x}(t)=\tr_A\left(R_{AB}\sqrt{M_{a\vert x}}\rho_A\sqrt{a\vert x}\right).
\end{equation}
If the state over time is separable, i.e. $R_{AB}=\sum_kp_k\rho_k^A\otimes\rho_k^B,$ we can express $\rho_{a\vert x}(t)$ as
\begin{equation}
    \rho_{a\vert x}(t)=\sum_k p_k\tr\left(\sqrt{M_{a\vert x}}\rho_k^A\sqrt{M_{a\vert x}}\right)\rho_k^B.
\end{equation}
State sent to Bob can be described as HSM, if the quantum mechanical description can be replaced with local response functions $D(a\vert x,k)$, i.e.,
\begin{equation}
    \tr\left(\sqrt{M_{a\vert x}}\rho_k^A\sqrt{M_{a\vert x}}\right)\to D(a\vert x,k).\label{eq:temporalHSM}
\end{equation}
Every separable state over time admits temporal HSM. Thus, inseparability of $R$ is a necessary condition for temporal steering under the NSIT assumption. Note that above considerations are true for all operators of the form (\ref{eq:star}) which satisfy $R_{AB}=\rho_A E_{B\vert A}$ for $\rho_A=\openone/d$.

\par However, numerical calculations show that for initial vacuum state $\vert 0\rangle$ undergoing phase damping a temporally steerable system may not be time-like entangled (see Fig.~\ref{fig:hierarchy_pdm}). Therefore, we conclude that temporal steering implies time-like entanglement, if NSIT condition is satisfied. If the NIST condition is violated, we ce can see in Fig.~\ref{fig:hierarchy_pdm} that the violation of hierarchy is possible.
 
\par A hierarchical relation between temporal steering and temporal Bell nonlocality holds for an arbitrary quantum system. Equation~(\ref{eq:temporalHSM}) yields the following probability distribution 
\begin{equation}
    p(a,b\vert x,y)=\sum_k p_k q(a\vert x,k)\tr\left(M_{b\vert y}\rho_k^B\right),\label{eq:temporalLHV}
\end{equation}
where $q(a\vert x,k) = \tr\left(\sqrt{M_{a\vert x}}\varrho_k^A\sqrt{M_{a\vert x}}\right)$ and $M_{b\vert y}$ denotes Bob's POVM at time $t_B$. The above distribution is an LHV model, if probabilities $q(a\vert x,k)$ and $\tr(M_{b\vert y}\varrho_k^B)$ are equivalent to $D(a\vert x,k)$ and $D(b\vert y,k),$ respectively. Therefore, every state over time admitting temporal LHV model is also temporally unsteerable. However, there can exist unsteerable states with no corresponding LHV model (i.e., nonlocal in time). Thus, the set of steerable states is a subset of nonlocal states.

\par To visualize relations between temporal correlations, we simulated the case of single qutrits initially in states $\openone/3$, $\vert 0\rangle$ and $\tfrac{1}{\sqrt{3}}(\vert 0\rangle+\vert 1\rangle+\vert 2\rangle)$, transmitted through standard quantum channels. Note that for initially pure states NSIT condition does not hold, which implies temporal steerability. The robustness-based measures of temporal entanglement, temporal steering, and temporal Bell nonlocality are shown in Fig.~\ref{fig:hierarchy_pdm}. Both TNR and TSR shown in Fig.~\ref{fig:hierarchy_pdm} were calculated for optimal PVMs (projection-valued measures) instead of POVMs, which were too computationally challenging to find in reasonable time. However, for a few points we found that the respective values of robustness for suboptimal POVMs are larger than for optimal PVMs. The depicted quantitative relations $\text{TER}\geq\text{TSR}\geq\text{TNR}$ and $\text{ER}\geq\text{TER}$ in segments (a), (d) and (g) illustrate to the hierarchy of correlations. Note that the same results as for POVMs hold by the same reasoning for PVMs (subsets of POVMs). 

\section{Conclusions}\label{sec:conclusions}
In this paper we have provided a comparison of $d$-dimensional temporal correlations. In order to standardize a description of different types of temporal correlations,  we have applied a consistent robustness-based measures and introduced \textit{temporal entanglement robustness} and \textit{temporal nonlocality robustness} as temporal counterparts of \textit{entanglement robustness} defined by Vidal and Tarrach in Ref.~\cite{PhysRevA.59.141} and \textit{nonlocality robustness} described in Ref.~\cite{PhysRevA.93.052112}. Additionally, we have proven that TER is a proper causality monotone. 

\par We have also extended the notion of PDO, firstly defined in multi-qubit systems~\cite{fitzsimons2015quantum}, to odd-dimensional (\ref{eq:new_PDO}). It seems that thought measurements in the PDO definition have to be non-contextual in order to make the PDO independent of a chosen basis. This might be connected with the formulation of the PDO in terms of decoherence functionals \cite{Zhang2020}. Expectation values of the product of the result of the thought measurements can be expressed in terms of diagonal decoherence functionals, which eliminate contextuality. 

\par We have numerically studied the robustness of the 
above-mentioned temporal correlations for a 3-level system initially in pure and mixed states with protective measurements. Our results indicate that in contrast to TSR, some level of quantum coherence is required for the initial state to exhibit nonzero TNR. As demonstrated in Fig.~\ref{fig:hierarchy_pdm} the standard hierarchy of correlations is conserved only in these cases, where the initial state satisfies NSIT condition. In the other cases (initially pure states) we can even witness the lack of temporal-entanglement and at the same time equality between nonzero values of TSR and TNR. The latter has deep physical meaning as this is numerical evidence of equivalence of TSR and TNR under certain conditions, possibly this equivalence could be not limited to pure states. It is also relevant to note that our analysis of temporal nonlocality (for zero evolution time) can be conceptually related to quantifying quantum contextuality with contextual fraction~\cite{Abramsky2017}.

\par A universal and sufficient condition for separability of a general $d$-dimensional system  is not known, thus, we could only use state-dependent criteria for separability in order to compare space-like and temporal entanglement. The complete characterization of the nontrivial relation between these two types of temporal correlations requires further research. However, we believe that our result will stimulate further research on defining temporal entanglement in a way that will not require the NSIT condition to be satisfied and will conserve the hierarchy of correlations known for spatially correlated systems.

\par In our study of the above-mentioned temporal correlations we have shown that for systems satisfying NSIT condition the hierarchy holds for every definition of a state over time which preserves the classical limit. We have also found an example of system violating both NSIT and the hierarchy between the correlations for two different definitions of a state over time. This answers the question formulated in the remarks of Ref.~\cite{PhysRevA.98.022104}. Relations between quantum temporal correlations do not depend on a particular definition of the star product.

\appendix
\section{Proof of TER being a causality monotone}\label{App:A}
An arbitrary PDO $R$ can be expressed as
\begin{equation}\label{eq:TERappend}
    R=(1+\gamma)\rho-\gamma\mathfrak{R}=0,
\end{equation}
where $\rho$ and $\mathfrak{R}$ are density and pseudo-density operators, respectively. TER is defined as the minimal $\gamma\geq 0$ which satisfies Eq.~(\ref{eq:TERappend}). If $R\geq0,$ then trivially $\textrm{TER}=0$. From a diagonal form of $R$, i.e. $R=\sum_{i=0}^{d^2-1}\mu_i\vert \mu_i\rangle\langle\mu_i\vert$, we see that $\gamma$ is minimal for $\gamma\mathfrak{R}=\sum_{i=0}^{n-1}\vert\mu_i\vert\vert \mu_i\rangle\langle\mu_i\vert$, where only first $n$ eigenvalues $\mu_i$ are negative. Notice that operator $\mathfrak{R}$ is positive semi-definite. It is clear that TER is maximal when $f(R)$ is maximal, viz for two consecutive measurements on a closed system. The evolution in a closed system is unitary.

\par The trace of $(1+\gamma)\rho$ is invariant under unitary operations, thus, the second condition for a causality monotone holds. Furthermore, Eq.~(\ref{eq:TERappend}) can be rewritten as
\begin{equation}
    \mathcal{E}\left(R\right)=\left(1+\gamma\right)\mathcal{E}\left(\rho\right)-\gamma \mathcal{E}\left(\mathfrak{R}\right),
\end{equation}
where $\mathcal{E}$ stands for a local, completely positive trace non-increasing map. Here $\gamma$ is not essentially optimal for state $\mathcal{E}(R)$. Thus, TER meets the third requirement for a causality monotone. 

\par Finally, TER can be considered a causality monotone if we prove that $\sum_kp_k\Phi(R_k)\geq \Phi(\sum_kp_kR_k).$ To this end let us consider an optimal $\gamma_1$,$\gamma_2$, such that
\begin{equation}
    R_k=\left(1+\gamma_k\right)\rho_k-\gamma_k\mathfrak{R}_k=0,\quad k\in\left\{1,2\right\}.
\end{equation}
For $R=pR_1+(1-p)R_2$ we obtain
\begin{eqnarray}
    R&=&p\left[\left(1+\gamma\right)\rho_1-\gamma\mathfrak{R}_1\right]+\left(1-p\right)\left[\left(1+\gamma\right)\rho_2-\gamma\mathfrak{R}_2\right]\nonumber\\
    &\equiv&\left(1+\gamma\right)\rho-\gamma\mathfrak{R},
\end{eqnarray}
where
\begin{eqnarray*}
    \rho=\frac{1+\gamma_1}{1+\gamma} p\rho_1&+&\frac{1+\gamma_2}{1+\gamma}\left(1-p\right)\rho_2,\\
    \mathfrak{R}=\frac{\gamma_1}{\gamma} p R_1&+&\frac{\gamma_2}{\gamma}\left(1-p\right)R_2,\\
    \gamma=p\gamma_1&+&\left(1-p\right)\gamma_2.
\end{eqnarray*}
By the definition $\textrm{TER}(pR_1+(1-p)R_2)\leq p\,\mathrm{TER}(R_1)+(1-p)\,\mathrm{TER}(R_2)$.

\section{SDP for calculating temporal nonlocality robustness}\label{App:B} 
Here we demonstrate that optimization problems (\ref{eq:TNR}) and (\ref{eq:TNR_SDP}) are equivalent. By requiring that behavior $Q(a,b\vert x,y)$  is a valid probability distribution, we can express definition (\ref{eq:TNR}) as
\begin{eqnarray}
    \text{TNR}&&\,=\min\,\beta\nonumber\\*
    \text{s.t. }&&\frac{P\left(a,b\vert x,y\right)+\beta Q\left(ab\vert xy\right)}{1+\beta}\nonumber\\*
    &&=\sum_{\mu,\nu}r_{\mu\nu}D\left(a\vert x,\mu\right)D\left(b\vert y,\nu\right),\nonumber\\*
    &&Q\left(a,b\vert x,y\right)\geq 0\quad\forall a,b,x,y,\nonumber\\*
    &&\sum_{a,b}Q\left(a,b\vert x,y\right)=\sum_{a,b}Q\left(a,b\vert x',y'\right)=1,\nonumber\\*
    &&r_{\mu\nu}\geq 0 \quad\forall\mu,\nu,\nonumber\\*
    &&\sum_{\mu\nu}r_{\mu\nu}=1,\quad\beta\geq 0,\label{eq:TNR_AppendixA}
\end{eqnarray}
where $r_{\mu\nu}\equiv r(\mu,\nu)$. Note that the behavior $Q(a,b\vert x,y)$ can be eliminated. Using the first constraint from (\ref{eq:TNR_AppendixA}) we obtain
\begin{eqnarray}
    Q\left(a,b\vert x,y\right)&=&\frac{1+\beta}{\beta}\sum_{\mu,\nu}r_{\mu\nu}D\left(a\vert x,\mu\right)D\left(b\vert y,\nu\right)\nonumber\\*
    &&-\frac{1}{\beta}P\left(a,b\vert x,y\right).\label{eq:Q(ab|xy)}
\end{eqnarray}
By requiring that all the distributions are normalized to the same value we impose another constraint:
\begin{equation}
    \sum_{x,y,a,b}P\left(a,b\vert x,y\right)=\sum_{x,y,a,b}\sum_{\mu,\nu}r_{\mu\nu}D\left(a\vert x,\mu\right)D\left(b\vert y,\nu\right).\label{eq:normalize_appendA}
\end{equation}

\par Let us introduce an unnormalized probability distribution of local variables $\tilde{r}_{\mu\nu}\equiv (1+\beta)r_{\mu\nu}$. Then, Eq.~(\ref{eq:normalize_appendA}) can be expressed as
\begin{eqnarray}
    (1+\beta)&&\,\sum_{x,y,a,b}P\left(a,b\vert x,y\right)\nonumber\\*
    =&&\,\sum_{x,y,a,b}\sum_{\mu,\nu}\tilde{r}_{\mu\nu}D\left(a\vert x,\mu\right)D\left(b\vert y,\nu\right).
\end{eqnarray}
Hence, TNR becomes
\begin{equation}
    \beta=\frac{\sum_{x,y,a,b,\mu,\nu}\tilde{r}_{\mu\nu}D\left(a\vert x,\mu\right)D\left(b\vert y,\nu\right)}{\sum_{x,y,a,b}P\left(a,b\vert x,y\right)}-1.
\end{equation}
For all $\mu$ and $\nu$ we have $\tilde{r}_{\mu\nu}\geq 0$, thus, one can rewrite the second condition from Eq.~(\ref{eq:TNR_AppendixA}) as
\begin{equation}
    \sum_{\mu,\nu}\tilde{r}_{\mu\nu}D\left(a\vert x,\mu\right)D\left(b\vert y,\nu\right)\geq P\left(a,b\vert x,y\right)\quad\forall a,b,x,y.
\end{equation}
Hence, we have eliminated all variables except for $\tilde{r}$. The above considerations allow expressing optimization problem (\ref{eq:TNR_AppendixA}) in following the SDP form,
\begin{eqnarray}
\text{TNR}=&&\,\min\left[\frac{\sum_{x,y,a,b,\mu,\nu}\tilde{r}_{\mu\nu}D(a\vert x,\mu)D(b\vert y,\nu)}{\sum_{x,y,a,b}P(a,b\vert x,y)}-1\right],\nonumber\\*
\text{s.t. }&&\sum_{\mu,\nu}\tilde{r}_{\mu\nu}D\left(a\vert x,\mu\right)D\left(b\vert y,\nu\right)\geq P\left(a,b\vert x,y\right),\nonumber\\*
 {}    &&\tilde{r}_{\mu\nu}\geq 0\quad\forall\mu,\nu,x,y,a,b.
\end{eqnarray}

\section{Separability conditions for a qutrit transmitted via standard quantum channels}\label{App:C}

\begin{figure*}
\includegraphics[width=\linewidth]{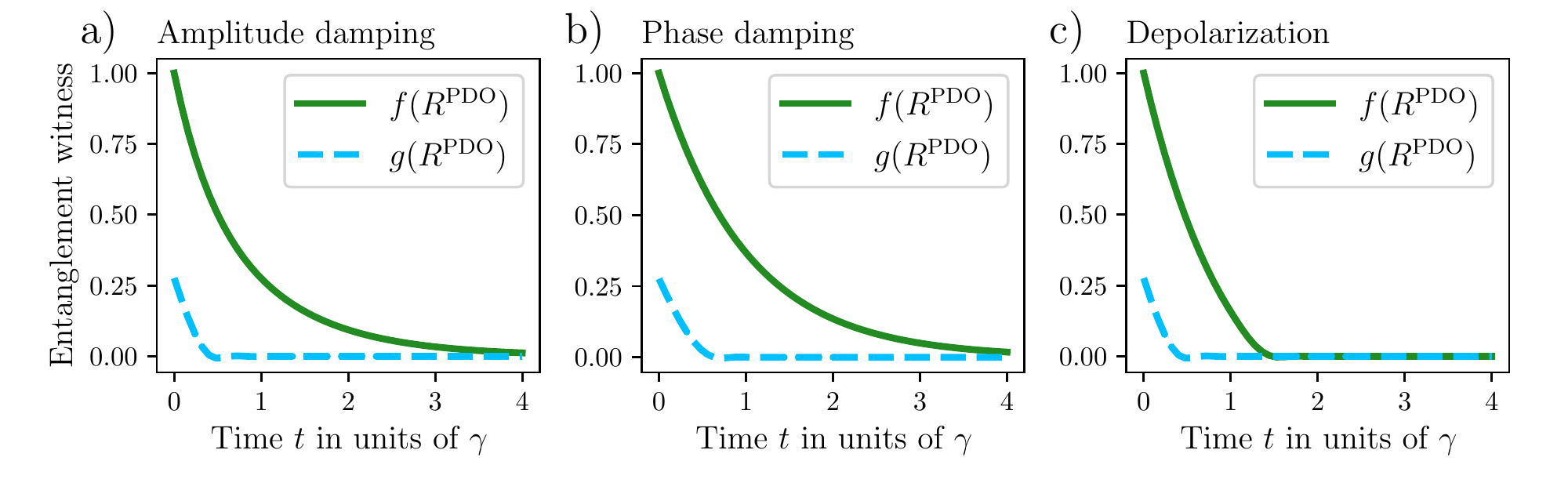}
\caption{(color online) The dynamics of $f$ and $g$ functions for a qutrit initially in maximally mixed state $\rho_A=\openone/3$, undergoing amplitude damping channel (a), phase damping channel (b), and depolarizing channel (c), respectively. The solid curves represent the dynamics of $f$ function and the dashed curves demonstrate the evolution of $g$. We can see that in all cases separability criterion (\ref{eq:separability}) is satisfied if a given PDO is positive semi-definite.}
\label{fig:separability}
\end{figure*}

\par Consider operator $\rho$ living in a product space $\mathcal{H}_A\otimes\mathcal{H}_B$, which can be written in terms of generators of special unitary group $\lambda_i$, normalized to $\tr(\lambda_i^2)=1$, as
\begin{equation}
    \rho=\sum_{i,j}\mathcal{C}_{ij}\lambda_i\otimes\lambda_j\label{eq:rhoBloch}
\end{equation}
and has positive semi-definite reduced states $\tr_A(\rho)$ and $\tr_B(\rho)$. Sarbicki \textit{et al.} in Ref.~\cite{PhysRevA.101.012341} showed that such $\rho$ is separable only if
\begin{equation}
    \|D_x^A\mathcal{C}D_y^B\|_{\tr}\leq\sqrt{\frac{(d_A-1+x^2)(d_B-1+y^2)}{d_Ad_B}},\label{eq:separability}
\end{equation}
where $D_x^A=\textrm{diag}\{x,1,\dots,1\}$, $D_y^B=\textrm{diag}\{y,1,\dots,1\}$ are $d_A^2\times d_A^2$ and $d_B^2\times d_B^2$ matrices, respectively, with arbitrary parameters $x,y$. Of course, $R^{\text{W}}$ and $R^{\text{PDO}}$ may be rewritten in terms of $\lambda_i$, despite definitions based on the phase-space operators. $\tr_X(R^{\text{PDO}}),\,\tr_X(R^{\text{PDO}})\geq 0$, where $X=A,B,$ so we may check space-like entanglement of a state over time under the action of an arbitrary quantum channel. 

\par The respective transformations of a single qutrit $\rho=\sum_{i,j}\rho_{ij}\vert i\rangle\langle j\vert$ under the action of an amplitude-damping, a phase-damping, and a depolarizing channel can be expressed as
\begin{widetext}
\begin{equation}
    \mathcal{E}_A(\rho)=
    \begin{bmatrix}
    e^{-2\gamma t}\rho_{00}&e^{-3\gamma t/2}\rho_{01}&e^{-\gamma t}\rho_{02}\\
    e^{-3\gamma t/2}\rho_{10}&2(e^{-\gamma t}-e^{-2\gamma t})\rho_{00}+e^{-\gamma t}\rho_{11}&\sqrt{2}(e^{-\gamma t/2}-e^{-3\gamma t/2})\rho_{01}+e^{-\gamma t/2}\rho_{12}\\
    e^{-\gamma t}\rho_{20}&\sqrt{2}(e^{-\gamma t/2}-e^{-3\gamma t/2})\rho_{10}+e^{-\gamma t/2}\rho_{21}&(e^{-2\gamma t}-2e^{-\gamma t}+1)\rho_{00}+(1-e^{-\gamma t})\rho_{11}+\rho_{22}
    \end{bmatrix},
\end{equation}
\begin{equation}
    \mathcal{E}_P(\rho)=
    \begin{bmatrix}
    \rho_{00}&e^{-\gamma t}\rho_{01}&e^{-\gamma t}\rho_{01}
    \\e^{-\gamma t}\rho_{10}&\rho_{11}&e^{-\gamma t}\rho_{12}
    \\e^{-\gamma t}\rho_{20}&e^{-\gamma t}\rho_{21}&\rho_{22}
    \end{bmatrix},
\end{equation}
\begin{equation}
    \mathcal{E}_D(\rho)=
    \begin{bmatrix}
    e^{-\gamma t}\rho_{00}+\frac{1}{3}(1-e^{-\gamma t})&e^{-\gamma t}\rho_{01}&e^{-\gamma t}\rho_{02}
    \\e^{-\gamma t}\rho_{10}&e^{-\gamma t}\rho_{11}+\frac{1}{3}(1-e^{-\gamma t})&e^{-\gamma t}\rho_{12}
    \\e^{-\gamma t}\rho_{20}&e^{-\gamma t}\rho_{21}&e^{-\gamma t}\rho_{22}+\frac{1}{3}(1-e^{-\gamma t})
    \end{bmatrix},
\end{equation}
\end{widetext}
where $\gamma$ stands for a decay rate. We can easily calculate a correlation tensor as $\mathcal{C}_{ij}=\tr[R(\lambda_i\otimes\lambda_j)]$. It turns out that in our case condition (\ref{eq:separability}) is the strongest for parameters $x=y=0$. Let us define a function $g(R)$ to quantify the above separability condition,
\begin{equation}
g(R)=
\begin{cases}
    \|D_0^A\mathcal{C}D_0^B\|_{\tr}-\frac{2}{3}\quad\text{for}\quad\|D_0^A\mathcal{C}D_0^B\|_{\tr}\geq \frac{2}{3}
    \\0\quad\text{for}\quad\|D_0^A\mathcal{C}D_0^B\|_{\tr}\leq\frac{2}{3}
\end{cases}.
\end{equation}
Fig.~\ref{fig:separability} presents dynamics of a normalized causality monotone $f(R^{\text{PDO}})$ and $g(R^{\text{PDO}})$ for an initial state $\rho_A=\openone/3$. We see that for three standard quantum channels inequality (\ref{eq:separability}) holds if a PDO $R^{\text{PDO}}$ is positive semi-definite. Also for W formulation of a state over time and initial states $\vert 0\rangle$ and $\tfrac{1}{\sqrt{3}}(\vert 0\rangle+\vert 1\rangle+\vert 2\rangle)$ condition $f(R)\geq 0$ is stronger than (\ref{eq:separability}).
\vspace*{5mm}
\section*{Acknowledgments} 
Authors acknowledge financial support by the Czech Science Foundation under the project No. 19-19002S and the financial support of the Polish National Science Center under grant No. DEC-2019/34/A/ST2/00081. 

\bibliography{main}

\end{document}